\documentclass[%
 reprint,
 superscriptaddress,
 amsmath,amssymb,
 aps,
]{revtex4-2}

\usepackage{stmaryrd}
\usepackage{graphicx}
\usepackage{dcolumn}
\usepackage{bm}
\usepackage{physics} 
\usepackage{braket}
\usepackage{hyperref}
\hypersetup{
  colorlinks   = true,    
  urlcolor     = blue,    
  linkcolor    = blue,    
  citecolor    = blue      
}
\usepackage[capitalize]{cleveref}
\usepackage{accents}
\usepackage{enumerate}
\usepackage[T1]{fontenc}
\usepackage{tabularx}

\DeclareDocumentCommand\ceil{ s m }
{
    \lceil #2 \rceil
}
\DeclareDocumentCommand\floor{ s m }
{
    \lfloor #2 \rfloor
}

\usepackage{xcolor}
\definecolor{myyellow}{HTML}{f1a226}
\definecolor{mygreen}{HTML}{298c8c}
\usepackage{graphicx}
\usepackage{placeins}

\begin{document}

\preprint{APS/123-QED}

\title{Experimental Characterization and Modeling of Measurement-Induced State-Transitions in a Fluxonium Superconducting Qubit}

\author{Martijn F. S. Zwanenburg}
\email[Contact author: m.f.s.zwanenburg@tudelft.nl]{}
\affiliation{QuTech and Kavli Institute of Nanoscience, Delft University of Technology, 2628 CJ, Delft, The Netherlands}
\author{Jinlun Hu}
\affiliation{QuTech and Kavli Institute of Nanoscience, Delft University of Technology, 2628 CJ, Delft, The Netherlands}
\author{Eugene Y. Huang}
\affiliation{QuTech and Kavli Institute of Nanoscience, Delft University of Technology, 2628 CJ, Delft, The Netherlands}
\author{Figen Yilmaz}
\affiliation{QuTech and Kavli Institute of Nanoscience, Delft University of Technology, 2628 CJ, Delft, The Netherlands}
\author{Siddharth Singh}
\affiliation{QuTech and Kavli Institute of Nanoscience, Delft University of Technology, 2628 CJ, Delft, The Netherlands}
\author{Christian Kraglund Andersen}
\email[Contact author: c.k.andersen@tudelft.nl]{}
\affiliation{QuTech and Kavli Institute of Nanoscience, Delft University of Technology, 2628 CJ, Delft, The Netherlands}

\date{\today}

\begin{abstract}
Superconducting qubits are most often measured using dispersive readout, which, ideally, implements a projective quantum non-demolition (QND) measurement. While a larger readout drive can increase the signal and, thus, reduce discrimination errors in the readout, strong microwave drives may also cause non-QND errors by driving the qubit to a state outside the computational subspace. In this work, we experimentally characterize measurement-induced state transitions (MIST) in a fluxonium qubit over its full external flux range. We further numerically calculate the MIST errors, and find that the theory accurately predicts eleven experimentally identified regions with increased MIST. In addition to transitions to higher fluxonium levels, we also find that, at certain flux points, MIST errors are dominated by transitions that include the transmission-line-like array modes of the fluxonium's superinductor. The excellent match between theory and experiment validates that the models accurately predict the occurrence of MIST in these systems, and further highlights the influence of array modes in fluxonium readout. 
\end{abstract}

\maketitle


\section{Introduction}
\label{sec:introduction}
Superconducting qubits are commonly read out through a resonator that is dispersively coupled to the qubit \cite{cqed_2004, blais_cqed}. Using dispersive readout, recent work has demonstrated readout fidelities approaching or even exceeding 99.9\% and readout durations around and below a hundred nanoseconds \cite{fpa_tmon, readout_nakamura, readout_devoret, iqm_999}. Additionally, dispersive readout has been characterized to be a quantum non-demolition (QND) measurement to a very high \mbox{degree \cite{readout_devoret, iqm_999}}. In dispersive readout, the superconducting qubit induces a state-dependent frequency shift of the resonator, such that the state of the qubit can be determined by probing the resonator response with a microwave tone. However, the dispersive coupling is not, conventionally, the native coupling between the qubit and the resonator. Instead, the dispersive coupling arises from a low-photon-number approximation of the Jaynes-Cummings Hamiltonian for the case of a large frequency detuning between the resonator and the qubit. Thus, the dispersive approximation may break down when a sufficiently large number of photons is added to the \mbox{resonator \cite{cqed_2004, nonlinear_disp_regime}}. In general, the break-down of the dispersive approximation may lead to a degradation of the readout fidelity and, more importantly, lead to non-QND \mbox{errors \cite{dynamics_tmon_ionization, nesterov_readout, dumas_mist}}. Non-QND errors are especially detrimental in the context of quantum error correction (QEC), where repeated syndrome extraction is needed to confidently decode the computational errors \cite{repeated_qed, qec_trapped_ions, surfacecode_eth, overcoming_leakage, surfacecode_google, qec_atoms}. While the behavior beyond the dispersive regime is generally involved to model, there has recently been a large focus on modeling and experimentally measuring the non-QND errors of superconducting qubits, where such errors are commonly referred to as measurement-induced state transitions (MIST) \cite{mist_beyond_rwa, mist_within_rwa, dynamics_tmon_ionization, dumas_mist, nesterov_readout, tmon_readout_nop_dependence, tmon_ro_zeno, devoret_repeated_measurements, mist_connolly, mist_ist, exc_dynamics_tmon, qnd_flux_pulse, fx_ro_array_modes, mist_fx_chapple, fluxonium_qnd_ro}. Most of this recent literature has focused on the transmon qubit \cite{transmon}, which is the most widely used superconducting qubit type. In this work, we are specifically interested in experimentally measuring and numerically modeling these errors for a fluxonium qubit \cite{fluxonium} which, in recent years, has become a key contender for scalable quantum computing showing long coherence times and high-fidelity operations \cite{millisecond_coherence, fluxonium_initialization, qnd_flux_pulse, circular_polarized_drive, mit_cz}. 

There are two notable differences when considering MIST in transmons and fluxoniums. In a fluxonium, the charge dispersion of all relevant levels is exponentially suppressed by the quadratic potential of its large superinductor, such that it does not suffer from charge-offset-dependent MIST that can occur in transmons \cite{tmon_readout_nop_dependence, dumas_mist, mist_within_rwa}. On the other hand, it has been shown in recent work that the array of Josephson junctions commonly used to realize the fluxonium's superinductor can host so-called array modes which can induce additional MIST \cite{mist_fx_chapple, fx_ro_array_modes}. Recent theoretical results suggest that MIST may be less prevalent in fluxoniums qubits than in transmon \mbox{qubits \cite{nesterov_readout, mist_fx_chapple, fx_ro_array_modes}}. Additionally, experimental work has also shown that spurious two-level systems may impact MIST in fluxonium qubits \cite{fx_ro_kou, qnd_flux_pulse, fluxonium_qnd_ro}.

In this work, we measure MIST over the full external flux range of a single fluxonium qubit, and specifically aim at experimentally verifying existing models for predicting the occurrence of MIST. After performing a basic device characterization in Sec. \ref{sec:device_characterization}, we experimentally characterize the MIST in \mbox{Sec. \ref{sec:qnd_errors}}. The experiment identifies eleven regions of interest where there is an increased level of MIST. While all regions arise from an avoided crossing in the dressed Hilbert space of the system, five are induced by one of the two lowest array modes of the fluxonium. These results further emphasize the role of array modes in readout of fluxonium qubits, and verify that existing models can be used to predict the occurrence of MIST and to inform future designs of fluxoniums. 

\section{Device Characterization}
\label{sec:device_characterization}
\begin{figure}[t]
\centering
\includegraphics[width=\linewidth]{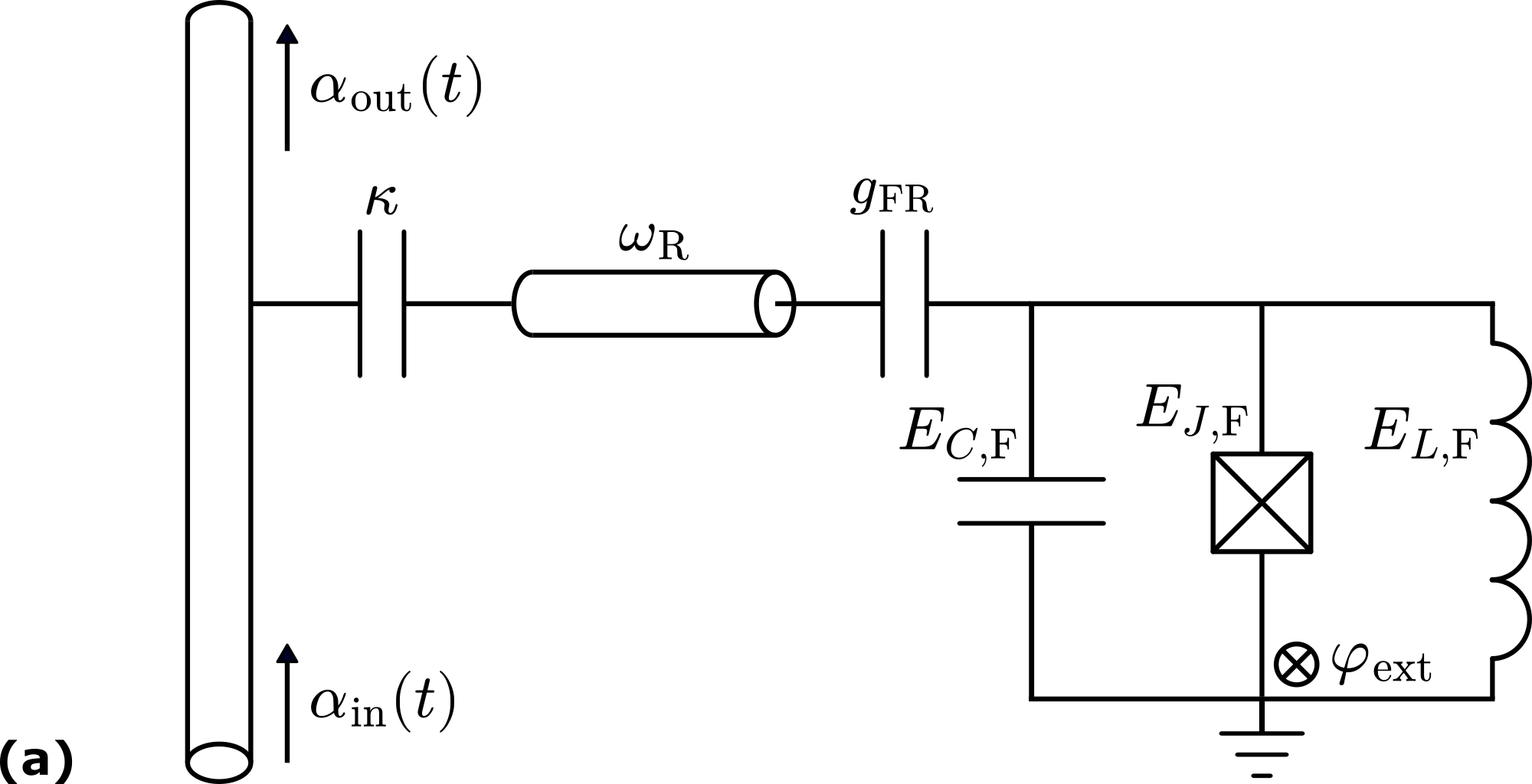}\\
\vspace{1em}
\includegraphics[width=\linewidth]{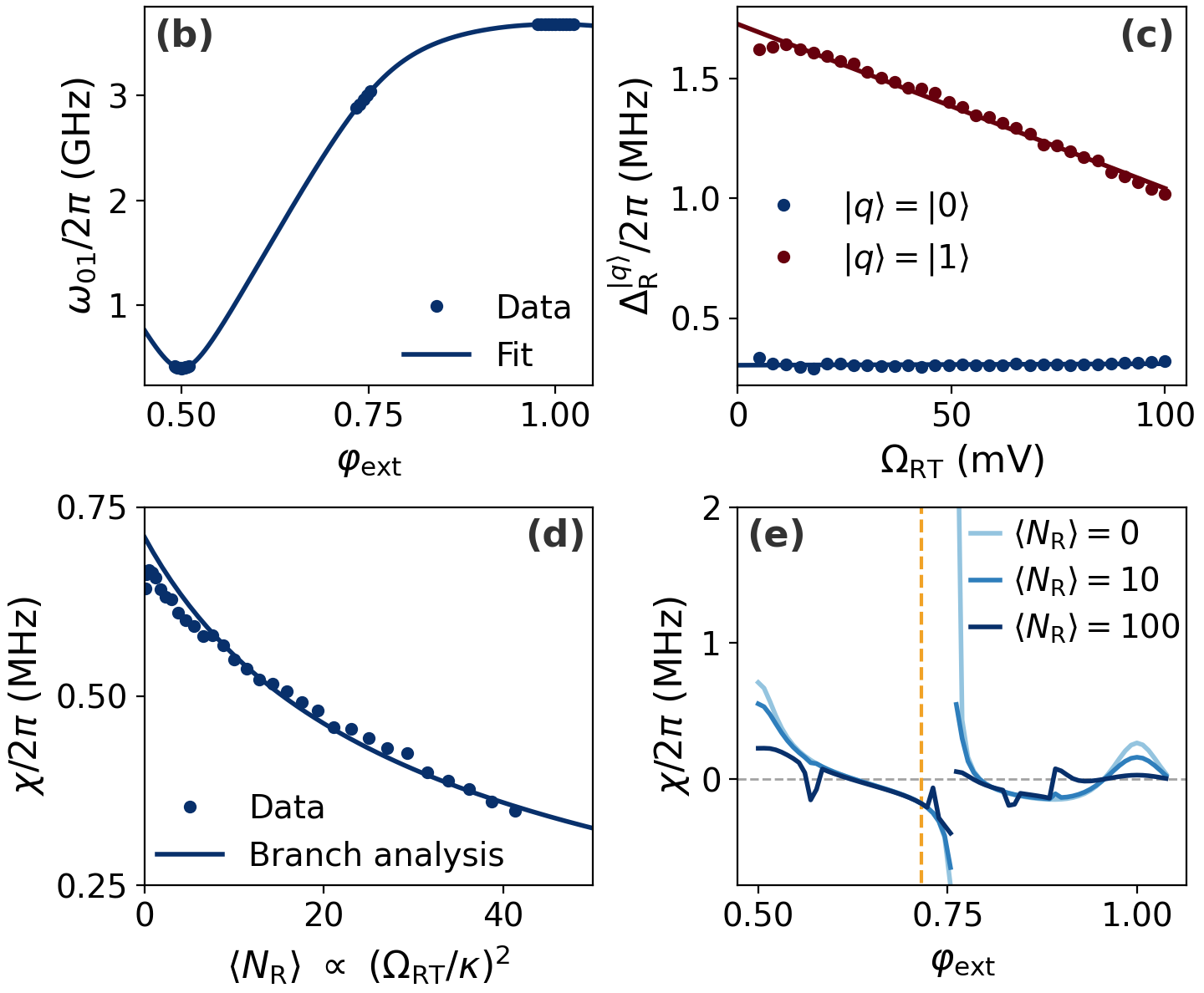}
\caption{\label{fig:figure1} (a) Circuit diagram of the fluxonium with a capacitively coupled readout resonator which, in turn, is capacitively coupled to a transmission line. (b) Two-tone spectroscopy data and fit used to extract the fluxonium's energy parameters. (c) Qubit-state-dependent resonator frequencies $\Delta_\text{R}^{\ket{q}} = \omega_\text{R}^{\ket{q}} - \omega_\text{R}$ at $\varphi_\text{ext}=0.5$ with the fluxonium prepared in $\ket{q}$ as a function of the room temperature (RT) readout amplitude $\Omega_\text{RT}$. The experimental data is linearly extrapolated (solid lines) to $\Omega_\text{RT}=0$ to obtain the vacuum resonator frequencies used to fit $\omega_\text{R}$ and $g_\text{FR}$. (d) Dispersive shift $\chi$ as a function of the drive strength expressed through the average photon number $\langle N_\text{R} \rangle$. The data is fitted with numerical results from branch analysis assuming a quadratic relation between $\langle N_\text{R} \rangle$ and $\Omega_\text{RT}$ to obtain the drive strength at the device. (e) Branch analysis simulations of the dispersive shift $\chi$ as a function of the external flux $\varphi_\text{ext}$ and the average photon number $\langle N_\text{R} \rangle$. The vertical, dashed line indicates the readout point used in later parts of the work. The horizontal, dashed line indicates $\chi=0$.}
\end{figure}

The experiments are performed on a single fluxonium schematically shown in Fig. \hyperref[fig:figure1]{\ref*{fig:figure1}(a)} together with the readout circuitry. The fluxonium additionally has a charge line used for microwave drive pulses and a flux line for applying static flux offsets and fast flux pulses, see also Appendix \ref{app:extended_data}. The fluxonium is described by the Hamiltonian
\begin{equation}
    \hat{H}_\text{F} = 4E_{C,\text{F}}\hat{n}^2 + \frac{1}{2}E_{L,\text{F}}\hat{\phi}^2 - E_{J,\text{F}}\cos(\hat{\phi} - 2\pi\varphi_\text{ext}),
\end{equation}
\noindent where $\hat{n}$ and $\hat{\phi}$ are the fluxonium's charge and phase operators, respectively, $\varphi_\text{ext}$ is the phase offset induced by the external flux bias, and $E_{C,\text{F}}$, $E_{J,\text{F}}$ and $E_{L,\text{F}}$ are its charging energy, Josephson energy, and inductive energy, respectively. The Hamiltonian of the readout resonator is given by $\hat{H}_\text{R} = \hbar\omega_\text{R}\hat{a}^\dagger\hat{a}$, where $\omega_\text{R}$ is its angular frequency and $\hat{a}$ and $\hat{a}^\dagger$ are its annihilation and creation operators, respectively. The linear inductor of the fluxonium is commonly realized by an array of $M$ Josephson junctions. As shown in recent work \cite{fx_ro_array_modes, mist_fx_chapple}, this array of Josephson junctions hosts (approximately) harmonic modes that can give rise to additional MIST. Therefore, we immediately include the lowest two array modes into the Hamiltonian, which are denoted as $\mu_1$ and $\mu_2$. We assume the modes are linear and further neglect direct coupling between the array modes. More details regarding these array modes are included in Appendix \ref{app:array_modes}. Similarly to the readout resonator, the individual array modes are described by the Hamiltonian $\hat{H}_{\mu_i} = \hbar\omega_{\mu_i}\hat{\mu}_i^\dagger\hat{\mu}_i$. Combining the fluxonium, resonator and the two array modes, and adding the input drive field and capacitive interactions, results in the full Hamiltonian
\begin{equation}
\label{eq:hamiltonian}
\begin{split}
    \hat{H} =& \hat{H}_\text{F} + \hat{H}_\text{R} + \hat{H}_{\mu_1} + \hat{H}_{\mu_2} \\
    & -i\hbar\sqrt{\kappa}\alpha_\text{in}(t)\left(\hat{a}-\hat{a}^\dagger\right) \\
    & -i\hbar g_\text{FR}\hat{n}\left(\hat{a} - \hat{a}^\dagger\right) \\
    & -i\hbar \hat{n}\sum_{i=1,2}\left( g_{\text{F}\mu_i}\left(\hat{\mu}_i - \hat{\mu}_i^\dagger\right)\right) \\
    & -\hbar \left(\hat{a} - \hat{a}^\dagger\right)\sum_{i=1,2}\left( g_{\text{R}\mu_i}\left(\hat{\mu}_i - \hat{\mu}_i^\dagger\right)\right).
\end{split}
\end{equation}
\noindent Here, the second line describes the input drive field acting on the resonator, where $\kappa$ is the resonator's linewidth and $\alpha_\text{in}(t) = \Omega\cos(\omega_dt)/\sqrt{\kappa}$ describes the classical input field with the drive strength $\Omega$ and drive frequency $\omega_d$. The classical output field shown in Fig. \hyperref[fig:figure1]{\ref*{fig:figure1}(a)} is related to the input field as $\alpha_\text{out}(t) = \alpha_\text{in}(t) - \sqrt{\kappa}\langle \hat{a}(t) \rangle$ \cite{gardiner_collett}. The third line describes the coupling between the fluxonium and the resonator with coupling strength $g_\text{FR}$.  The fourth line describes the coupling between the fluxonium and the array modes with coupling strength $g_{\text{F}\mu_i}$. The last line describes the coupling between the resonator and the array modes with coupling strength $g_{\text{R}\mu_i}$.

The energy parameters of the full system are determined through a series of experiments and simulations, and are summarized in Table \ref{tab:eparams}. First, the fluxonium's qubit frequency is measured as a function of the external flux with two-tone spectroscopy, see Fig. \hyperref[fig:figure1]{\ref*{fig:figure1}(b)}. The fluxonium's energy parameters are fitted to these experimental results while omitting the resonator and array modes. Next, the fluxonium is biased to its flux symmetry point $\varphi_\text{ext}=0.5$, where the qubit-state-dependent resonator frequency $\omega_\text{R}^{\ket{q}}$ is measured spectroscopically as a function of the room temperature (RT) readout strength $\Omega_\text{RT}$. The results are shown in Fig. \hyperref[fig:figure1]{\ref*{fig:figure1}(c)}, which show a linear dependence of the resonator frequencies on the drive strength. The resonator frequencies are linearly extrapolated to zero drive strength to obtain the qubit-state dependent resonator frequency for $\langle N_\text{R} \rangle = \langle \hat{a}^\dagger\hat{a}\rangle = 0$. We now proceed in one of two ways. In the first approach, the array modes are neglected, and $\omega_r$ and $g_\text{FR}$ are fitted directly to the qubit-state-dependent resonator frequencies. These fitted parameters are shown in the top half of Table \ref{tab:eparams}. In the second approach we include the lowest two array modes. We experimentally obtain the frequency of the first array mode with two-tone spectroscopy, and implement the fitting procedure detailed in Appendix~\ref{app:array_modes} to obtain the fitted parameters shown in the bottom half of Table \ref{tab:eparams} \cite{fx_ro_array_modes, mist_fx_chapple}. Notably, the fitted $g_\text{FR}$ is approximately 20\% larger in the model that includes the array modes even though the effective dispersive shift is identical in the two fits, which indicates that the array modes decrease the effective fluxonium-resonator coupling.

\begin{table}[b]
\caption{\label{tab:eparams} Effective energy parameters of the systems without and with the array modes, see additional explanations in the main text.}
\begin{ruledtabular}
\begin{tabular}{llll}
\multicolumn{2}{c}{Systems (GHz)} & \multicolumn{2}{c}{Couplings (MHz)} \\
\hline
\multicolumn{4}{c}{Fitted without array modes} \\
$E_{J,\text{F}}/h$ & 2.571 & $g_\text{FR}/2\pi$ & 76.65 \\
$E_{C,\text{F}}/h$ & 0.872 & $\omega_\text{R}/2\pi$ & 6.7905 \\
$E_{L,\text{F}}/h$ & 0.449 & $\kappa/2\pi$ & 0.230  \\
\hline
\multicolumn{4}{c}{Fitted with array modes} \\
$E_{J,\text{F}}/h$ & 2.571 & $g_\text{FR}/2\pi$ & 91.71 \\
$E_{C,\text{F}}/h$ & 0.872 & $g_{\text{F}\mu_1}/2\pi$ & 19.67 \\
$E_{L,\text{F}}/h$ & 0.449 & $g_{\text{F}\mu_2}/2\pi$ & 807.40 \\
$\omega_\text{R}/2\pi$ & 6.7915 & $g_{\text{R}\mu_1}/2\pi$ & 33.27 \\
$\omega_{\mu_1}/2\pi$ & 8.86 & $g_{\text{R}\mu_2}/2\pi$ & 10.61 \\
$\omega_{\mu_2}/2\pi$ & 14.76 & $\kappa/2\pi$ & 0.230 \\
\end{tabular}
\end{ruledtabular}
\end{table}

To compute the photon-number-occupation of the readout resonator, the room temperature readout drive strength $\Omega_\text{RT}$ needs to be converted to the drive strength at the device $\Omega$. To compute this scaling factor, the photon-number-dependent dispersive shift is computed numerically with branch analysis \cite{dynamics_tmon_ionization, dumas_mist} for the system that does not include the array modes, and is fitted to the experimentally obtained dispersive shift in Fig.  \hyperref[fig:figure1]{\ref*{fig:figure1}(d)}. Specifically, we assume a linear dependence between $\Omega_\text{RT}$ and $\Omega$ and further assume that the resonator is in equilibrium, such that the average photon number is given by $\langle N_\text{R} \rangle = (\Omega/\kappa)^2 = (\beta\Omega_\text{RT}/\kappa)^2$, with $\beta$ the to-be-fitted scaling factor. In the experiment, we ensure the resonator is in equilibrium by performing the readout acquisition for only the final 2 $\mu$s of a 10 $\mu$s $\approx$ $14.5/\kappa$ readout pulse. Finally, Fig.  \hyperref[fig:figure1]{\ref*{fig:figure1}(e)} shows the simulated dispersive shift $\chi$ for varying average photon number and as a function of the external flux $\varphi_\text{ext}$ also obtained from branch analysis without the array modes. The vertical, dashed line in Fig.  \hyperref[fig:figure1]{\ref*{fig:figure1}(e)} indicates the readout point used in later parts of the work. 
\begin{figure}[b]
\centering
\includegraphics[width=\linewidth]{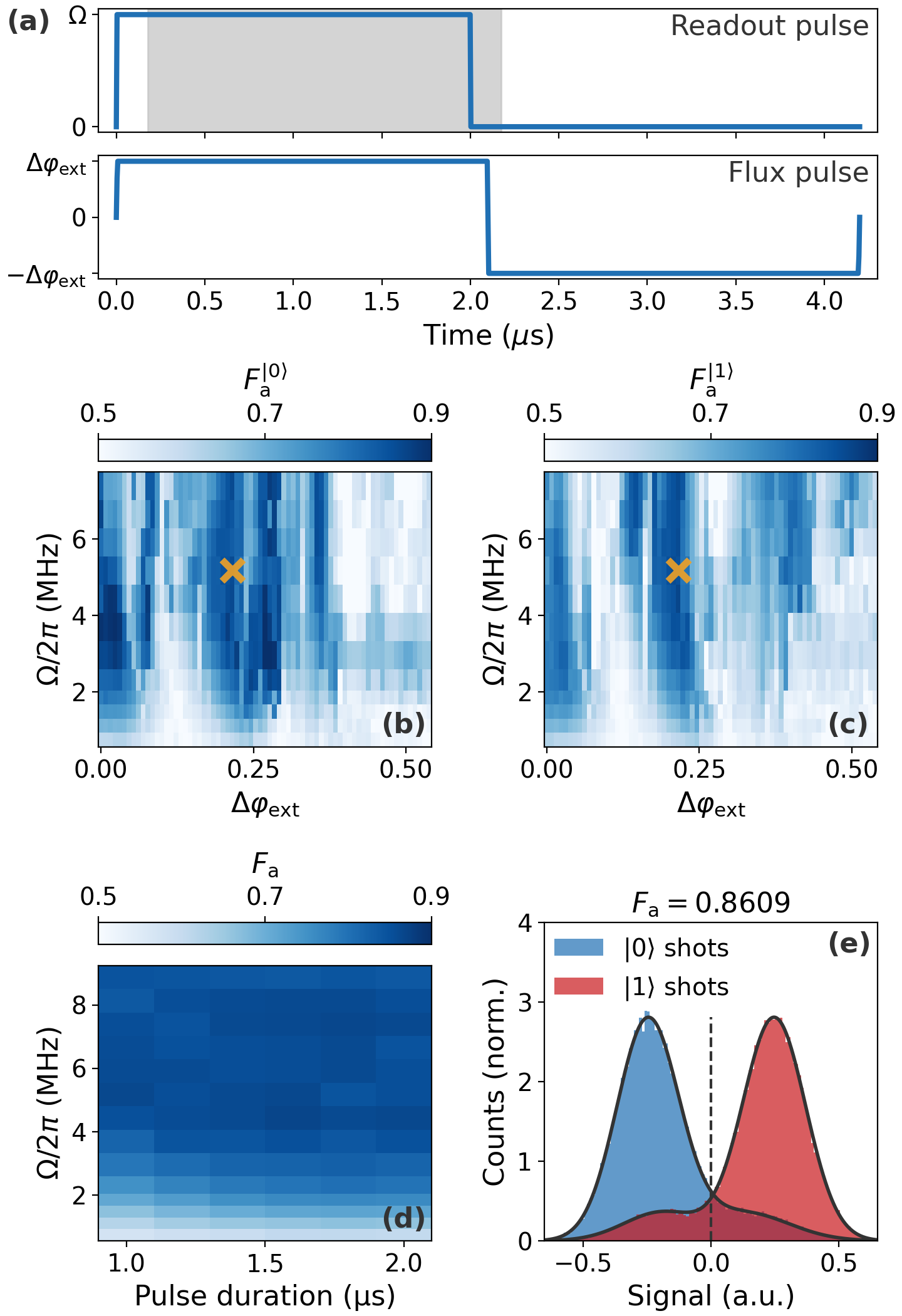}\\
\caption{\label{fig:figure3} (a) Unmodulated readout pulse and the net-zero flux pulse used for these experiments. The shaded area indicates the readout acquisition window. (b), (c) Assignment fidelity $F_\text{a}^{\ket{q}}$ of the $\ket{0}$ and $\ket{1}$ state, respectively, as a function of the flux pulse amplitude $\Delta\varphi_\text{ext}$ and the readout drive strength. The qubit is initialized at the sweet spot $\varphi_\text{ext}=0.5$, and tuned to the respective external fluxes with a fast flux pulse for readout. The marked point indicates the readout parameters used in panels (d) and (e) and in later parts of the work. \mbox{(d) Assignment} fidelity as a function of the readout pulse duration and drive strength. The integration time is fixed at \mbox{2 $\mu$s}. (e) Single-shot readout histograms of the $\ket{0}$ and $\ket{1}$ states. The solid lines indicate double-Gaussian fits, and the vertical, dashed line indicates the threshold used to compute the assignment fidelity.}
\end{figure}

After fitting the energy parameters, the readout assignment fidelity $F_\text{a}$ is characterized over the full external flux range of the fluxonium using the pulse sequence shown in \mbox{Fig. \hyperref[fig:figure3]{\ref*{fig:figure3}(a)}}, in which a readout pulse with drive strength $\Omega$ and a flux pulse that offsets the fluxonium from its symmetry point by $\Delta\varphi_\text{ext}$ are applied simultaneously. The gray, shaded area in \mbox{Fig. \hyperref[fig:figure3]{\ref*{fig:figure3}(a)}} indicates the integration window, which is offset from the start of the readout pulse to account for the time required for the signal to travel from the room-temperature electronics to the device and back \cite{latency}. The net-zero flux pulse \cite{net_zero} consists of a positive and negative ramp with a total duration of 4.2 $\mu$s to ensure that the negative ramp occurs only after the readout acquisition is complete. Prior to these readout pulses, the qubit is reset by a 20 $\mu$s flux-pulse to \mbox{$\varphi_\text{ext}=0.823$ \cite{fluxonium_initialization}}, and tuned back to $\varphi_\text{ext}=0.5$ for an optional $X_\pi$ pulse to prepare the qubit in $\ket{1}$ (not shown schematically). Figs. \hyperref[fig:figure3]{\ref*{fig:figure3}(b),(c)} show the qubit-state-dependent assignment fidelity $F_\text{a}^{\ket{q}}$ as a function of the external flux and readout drive strength. Prior to computing the assignment fidelities, the optimal integration weights are computed at each individual drive strength and flux bias \cite{bultink, heinsoo}, see \mbox{Appendix \ref{app:extended_data}}. As expected, $F_\text{a}^{\ket{q}}$ generally improves when the readout amplitude is increased. However, we see multiple flux bias regions with non-trivial behavior as well. For example, for very large readout drive amplitudes, the readout quality is generally worse and, for specific flux biases, the readout assignment fidelity is low independently of the drive strength. Only two flux bias regions feature a high average assignment fidelity $F_\text{a} = (F_\text{a}^{\ket{0}} + F_\text{a}^{\ket{1}})/2$, which are around the flux symmetry point and around $\Delta\varphi_\text{ext}=0.21$. The highest assignment fidelity is 86.1\%, which is reached at the marked point in \mbox{Figs. \hyperref[fig:figure3]{\ref*{fig:figure3}(b),(c)}} and is an improvement compared to the 82.9\% fidelity achieved at the flux symmetry point \cite{fpa_theory, fpa_experimental}. In \mbox{Fig. \hyperref[fig:figure3]{\ref*{fig:figure3}(d)}} we investigate the influence of the drive strength and readout pulse duration on the assignment fidelity $F_\text{a}$ at the optimal readout point while keeping the integration time fixed at 2 $\mu$s. The results show that shortening the readout drive pulse to \mbox{1 $\mu$s} has a negligible influence on the assignment fidelity, which is a result of the relatively small linewidth of the resonator. Increasing the readout drive strength beyond $\sim 5$ MHz does not improve the assignment fidelity which we ascribe to the onset of MIST. In the next section the readout is performed at the optimal readout point with a \mbox{1 $\mu$s} readout drive pulse and with $\Omega/2\pi = 5.18$ MHz, for which the single-shot readout histogram is shown in \mbox{Fig. \hyperref[fig:figure3]{\ref*{fig:figure3}(e)}}.

\section{Characterization of MIST}
\label{sec:qnd_errors}
\begin{figure*}[t]
\centering
\includegraphics[width=.75\linewidth]{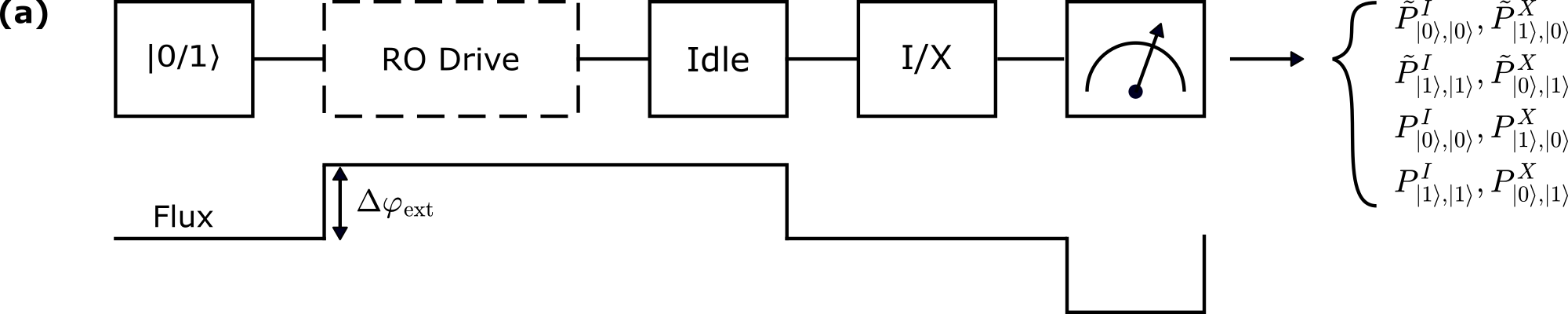} \\
\vspace{.1em}
\includegraphics[width=\linewidth]{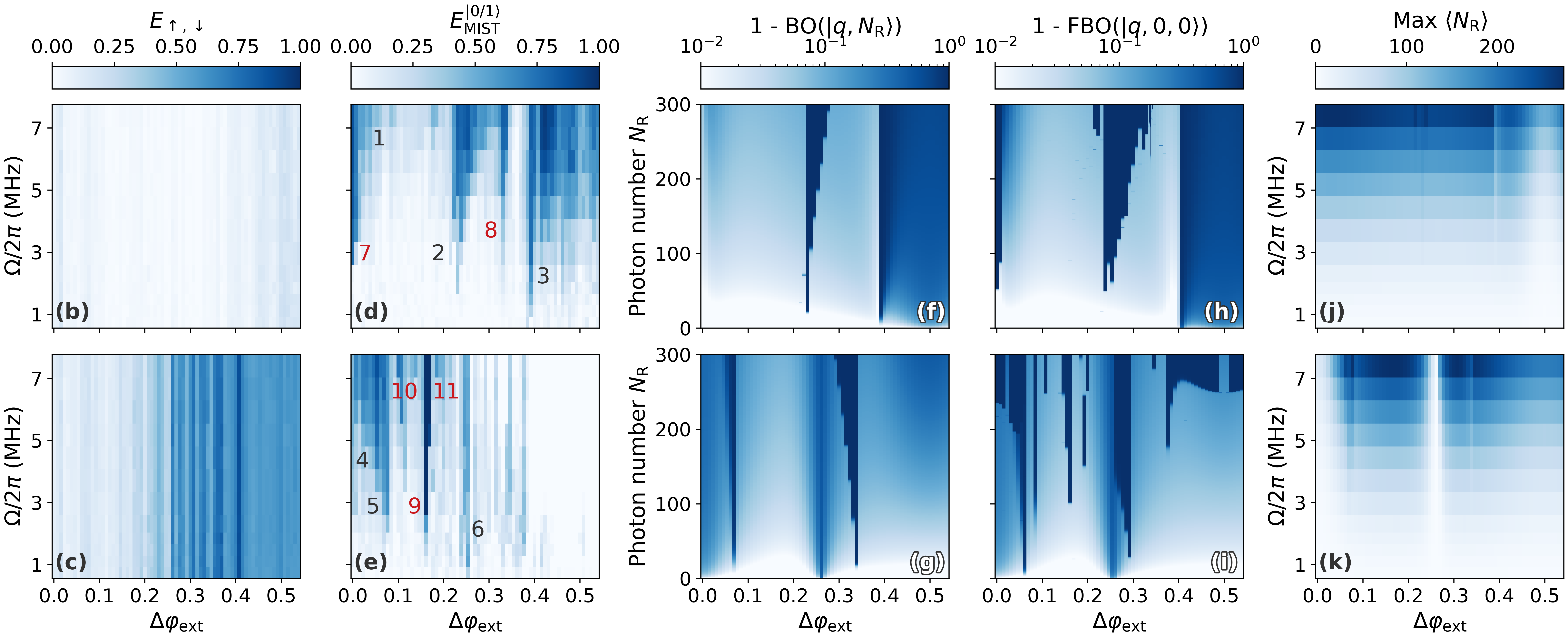}
\caption{\label{fig:figure4} (a) Circuit diagram of the experiment used to characterize MIST. The qubit is initialized in $\ket{0}$ or $\ket{1}$, and tuned away from sweet spot by $\Delta \varphi_\text{ext}$ with a fast flux pulse. Next, an optional readout drive pulse is played, followed by an idle duration to account for the time required for the resonator to deplete. The qubit is readout at the flux-point indicated in \mbox{Figs. \hyperref[fig:figure3]{\ref*{fig:figure3}(b),(c)}} after an optional $X$ operation performed at sweet spot. (b)-(k) Numerical and experimental results for a 1 $\mu$s readout drive pulse as a function of the flux pulse amplitude $\Delta \varphi_\text{ext}$ and with the qubit initialized in $\ket{0}$ (panels (b), (d), (f), (h) and (j)) or in $\ket{1}$ (panels (c), (e), (g), (i) and (k)). (b), (c) Experimentally obtained excitation error $E_\uparrow$ and relaxation error $E_\downarrow$, respectively. (d), (e) Experimentally obtained MIST error $E_\text{MIST}^{\ket{0/1}}$. The numbers annotate eleven regions with increased MIST of which the origin is detailed in the main text. (f), (g) Branch analysis results without the array modes. Specifically, we plot the complement of the branch overlap (BO) as defined in the main text. (h), (i) Complement of the Floquet branch overlap (FBO) computed with Floquet branch analysis including the lowest two array modes. (j), (k) Numerically computed maximum average photon number occupation of the readout resonator during the \mbox{1 $\mu$s} readout drive pulse.}
\end{figure*}

In our experiment, MIST cannot be measured directly and we will instead derive contributions to MIST from a characterization of the total non-QND errors in the system. To accomplish the separation of error sources in the measurement of non-QND errors, we use the experimental sequence shown in \mbox{Fig. \hyperref[fig:figure4]{\ref*{fig:figure4}(a)}}. First, the qubit is initialized in $\ket{0}$ or in $\ket{1}$. Next, an optional readout drive pulse is applied simultaneously with a fast flux pulse which offsets the fluxonium from its symmetry point. If no readout drive pulse is applied, the fluxonium idles at the specified flux point for the same duration as the readout drive pulse. Here, the duration of the optional readout drive pulse is fixed at 1 $\mu$s, and \mbox{Appendix \ref{app:extended_data}} shows the results for 0.5 $\mu$s and 2 $\mu$s durations. After the optional readout pulse, the fluxonium idles at the specified flux point for 3 $\mu$s $\approx 4.3/\kappa$ to ensure the resonator is depleted before flux-tuning back to the symmetry point. Here, either an identity or $X$ gate is performed prior to a measurement. The final measurement is always performed at the flux point highlighted in \mbox{Figs. \hyperref[fig:figure3]{\ref*{fig:figure3}(b),(c)}}, and with a fixed pulse duration, integration time, and drive strength, of which the single-shot readout histogram is shown in \mbox{Fig. \hyperref[fig:figure3]{\ref*{fig:figure3}(e)}}. After the measurement, an additional flux pulse is applied (not shown) to ensure the integrated flux pulse is net-zero \cite{net_zero}. Performing the sequence with and without the readout pulse allows for the distinction between MIST errors and other sources of non-QND errors such as flux-bias-dependent relaxation and thermal excitations. Furthermore, determining whether the populations swap between the $\ket{0} \leftrightarrow \ket{1}$ states upon performing the final $X$ gate allows for the detection of leakage to non-computational states. 

First, the sequence is performed without the readout drive pulse to obtain the probabilities $\tilde{P}^{I/X}_{\ket{i},\ket{j}}$, which is the probability that the qubit is measured in state $\ket{i}$ provided it was initialized in $\ket{j}$. We can express these probabilities using a simple model as a function of the assignment fidelities for each state $F_\text{a}^{\ket{q}}$ and $E_\uparrow$ and $E_\downarrow$ referring to the readout errors from thermal excitation or relaxation during the time between the initialization and the final readout, respectively. Thus, we arrive at the following probabilities,
\begin{align}
    & \tilde{P}^{I}_{\ket{0},\ket{0}} = F_\text{a}^{\ket{0}}\Big(1-E_\uparrow\Big) + \Big(1-F_\text{a}^{\ket{1}}\Big)E_\uparrow, \\
    & \tilde{P}^{X}_{\ket{1},\ket{0}} = F_\text{a}^{\ket{1}}\Big(1-E_\uparrow\Big) + \Big(1-F_\text{a}^{\ket{0}}\Big)E_\uparrow, \\
    & \tilde{P}^{I}_{\ket{1},\ket{1}} = F_\text{a}^{\ket{1}}\Big(1-E_\downarrow\Big) + \Big(1-F_\text{a}^{\ket{0}}\Big)E_\downarrow, \\
    & \tilde{P}^{X}_{\ket{0},\ket{1}} = F_\text{a}^{\ket{0}}\Big(1-E_\downarrow\Big) + \Big(1-F_\text{a}^{\ket{1}}\Big)E_\downarrow.
\end{align}
\noindent Using these expressions, the empirical errors $E_\uparrow$ and $E_\downarrow$ are calculated as
\begin{align}
    & E_\uparrow = \frac{F_\text{a} - \frac{\tilde{P}^{I}_{\ket{0},\ket{0}} + \tilde{P}^{X}_{\ket{1},\ket{0}}}{2}}{2F_\text{a} - 1},\\ 
    & E_\downarrow = \frac{F_\text{a} - \frac{\tilde{P}^{I}_{\ket{1},\ket{1}} + \tilde{P}^{X}_{\ket{0},\ket{1}}}{2}}{2F_\text{a} - 1}.
\end{align}
\noindent Next, the experiment is performed with the additional readout drive pulse to obtain the probabilities $P^{I/X}_{\ket{i},\ket{j}}$. We can similarly find a simple model for these probabilities as
\begin{align}
    & P^I_{\ket{0},\ket{0}} = \tilde{P}^I_{\ket{0},\ket{0}} + E_\text{MIST}^{\ket{0}} \left( P_{\ket{0},\ket{x}} - F_\text{a}^{\ket{0}} \right), \\
    & P^X_{\ket{1},\ket{0}} = \tilde{P}^X_{\ket{1},\ket{0}} + E_\text{MIST}^{\ket{0}}\left(P_{\ket{1},\ket{x}} - F_\text{a}^{\ket{0}}\right), \\
    & P^I_{\ket{1},\ket{1}} = \tilde{P}^I_{\ket{1},\ket{1}} + E_\text{MIST}^{\ket{1}}\left(P_{\ket{1},\ket{x}} - F_\text{a}^{\ket{1}}\right), \\
    & P^X_{\ket{0},\ket{1}} = \tilde{P}^X_{\ket{0},\ket{1}} + E_\text{MIST}^{\ket{1}}\left(P_{\ket{0},\ket{x}}- F_\text{a}^{\ket{0}}\right).
\end{align}
\noindent In this model, we include the possibility that the readout drive pulse excites the fluxonium to some unknown state $\ket{x}$ (or multiple states) with probability $E_\text{MIST}^{\ket{q}}$, with $E_\text{MIST}^{\ket{q}}$ the qubit-state-dependent MIST error. $P_{\ket{q},\ket{x}}$ denotes the probability that the measurement assigns this state $\ket{x}$ as $\ket{q}$. From here, $E_\text{MIST}^{\ket{q}}$ is calculated as
\begin{align}
    & E_\text{MIST}^{\ket{0}} = \frac{P^I_{\ket{0},\ket{0}} + P^X_{\ket{1},\ket{0}} - 2F_\text{a}}{1 - 2F_\text{a}} - 2E_\uparrow, \\
    & E_\text{MIST}^{\ket{1}} = \frac{P^I_{\ket{1},\ket{1}} + P^X_{\ket{0},\ket{1}} - 2F_\text{a}}{1 - 2F_\text{a}} - 2E_\downarrow.
\end{align}
We perform the pulse sequence in Fig. \hyperref[fig:figure4]{\ref*{fig:figure4}(a)} as a function of the flux-pulse amplitude $\Delta\varphi_\text{ext}$ and the drive strength $\Omega$ to obtain $E_\uparrow$, $E_\downarrow$, $E_\text{MIST}^{\ket{0}}$ and $E_\text{MIST}^{\ket{1}}$, see \mbox{Figs. \hyperref[fig:figure4]{\ref*{fig:figure4}(b)-(e)}}. We reiterate that the drive strength of the readout pulse for the final measurement remains fixed. While the thermal excitation errors shown in Fig. \hyperref[fig:figure4]{\ref*{fig:figure4}(b)} remain low for all parameters, the relaxation error in \mbox{Fig. \hyperref[fig:figure4]{\ref*{fig:figure4}(c)}} increases strongly for $\varphi_\text{ext}>0.75$. As shown in Fig. \hyperref[fig:extended_data]{\ref*{fig:extended_data}(b)} in \mbox{Appendix \ref{app:extended_data}}, the $T_1$ of the qubit drops to only a few microseconds in this regime, which severely limits ability to confidently extract $E_\text{MIST}^{\ket{1}}$. From the extracted $E_\text{MIST}^{\ket{q}}$, we observe eleven distinct regions with increased MIST as labeled in Figs. \hyperref[fig:figure4]{\ref*{fig:figure4}(d),(e)}. To identify their respective origins, we perform conventional branch analysis \cite{dynamics_tmon_ionization, dumas_mist} for the system that does not include the array modes with the fitted parameters listed in the top half of \mbox{Table \ref{tab:eparams}}. In these simulations, the fluxonium and resonator are truncated to 16 and 410 levels, respectively. To calculate an accurate proxy for the $E_\text{MIST}^{\ket{q}}$, which prevents simulating the full readout dynamics, we calculate the complement of the branch overlap (BO), which is defined as the overlap between $\ket{\overline{q,N_\text{R}}}$ and all states in the bare branch belonging to the fluxonium's $\ket{q}$ state
\begin{equation}
    \text{BO}(\ket{q,N_\text{R}}) = \sum_{N_\text{R}'} \left\vert \langle\overline{q,N_\text{R}} \vert q,N_\text{R}'\rangle \right\vert^2,
\end{equation}
\noindent where $\ket{\overline{\psi}}$ and $\ket{\psi}$ denote the dressed and bare eigenstates, respectively. These numerical simulations, see \mbox{Figs. \hyperref[fig:figure4]{\ref*{fig:figure4}(f),(g)}}, identify the following transitions as dominating in each region: (1) $\ket{0,N_\text{R}} \leftrightarrow \ket{3,N_\text{R}-1}$, (2) $\ket{0,N_\text{R},} \leftrightarrow \ket{15,N_\text{R}-4}$, (3) $\ket{0,N_\text{R}} \leftrightarrow \ket{3,N_\text{R}-1}$, (4) $\ket{1,N_\text{R}} \leftrightarrow \ket{4,N_\text{R}-1}$, (5) $\ket{1,N_\text{R}} \leftrightarrow \ket{8,N_\text{R}-2}$ and (6) $\ket{1,N_\text{R}} \leftrightarrow \ket{5,N_\text{R}-1}$. This conventional branch analysis leaves six clear regions unexplained including the very pronounced MIST observed for the $\ket{0}$ state at the flux symmetry point $\varphi_\text{ext} =0.5$. To determine whether they originate from the array modes of the fluxonium, we perform Floquet branch analysis for the system including the lowest two array modes with the parameters listed in the bottom half of Table \ref{tab:eparams}. In these simulations, the resonator is assumed to be in a coherent state, and the system is transformed to the displaced frame where the resonator is dropped completely \cite{classical_chaos, dumas_mist, fx_ro_array_modes, mist_fx_chapple}. The fluxonium is again truncated at 16 levels and the array modes are truncated at 5 levels. In the Floquet branch analysis implementation, we propagate the relevant branches using the Lanczos algorithm \cite{lanczos, dmrgx_google, crosstalk_zwanenburg}, and only diagonalize the full Hamiltonian when Lanczos does not converge. We now plot the complement of the Floquet branch overlap (FBO)
\begin{equation}
    \text{FBO}(\ket{q,0,0}) = \sum_{\mu_1',\mu_2'} \left\vert \langle{\overline{q,0,0}}\vert q,\mu_1',\mu_2' \rangle \right\vert^2,
\end{equation}
\noindent where $\ket{\overline{q,0,0}}$ are the dressed Floquet eigenstates in the displaced frame. The results of these simulations, shown in \mbox{Figs. \hyperref[fig:figure4]{\ref*{fig:figure4}(h),(i)}}, identify the dominating transitions for all five remaining regions with increased MIST: (7) $\ket{0,N_\text{R},0,0} \leftrightarrow \ket{3,N_\text{R}-2,1,0}$ and (8) $\ket{0,N_\text{R},0,0} \leftrightarrow \ket{3,N_\text{R}-2,1,0}$, (9) $\ket{1,N_\text{R},0,0} \leftrightarrow \ket{5,N_\text{R}-3,0,1}$, (10) $\ket{1,N_\text{R},0,0} \leftrightarrow \ket{7,N_\text{R}-3,0,1}$ and (11) $\ket{1,N_\text{R},0,0} \leftrightarrow \ket{8,N_\text{R}-4,0,1}$.

While the numerics show excellent quantitative agreement with the experimental data, there are also several minor differences. Firstly, the simulations shown in \mbox{Fig. \hyperref[fig:figure4]{\ref*{fig:figure4}(h)}} predict that region 8 should be very narrow, while it is significantly wider in the experimental data. We ascribe this discrepancy to small variations in $\Delta\varphi_\text{ext}$ during the readout drive pulse that arise from flux distortions \cite{cryoscope, hellings}. Secondly, there is a small deviation between the flux amplitudes $\Delta\varphi_\text{ext}$ of the regions with increased MIST found in the experiments and predicted by the branch analysis simulations with the array modes. For instance, region 3 appears just below $\Delta\varphi_\text{ext}=0.4$ in the experimental data in \mbox{Fig. \hyperref[fig:figure4]{\ref*{fig:figure4}(d)}}, while branch analysis with the array modes in \mbox{Fig. \hyperref[fig:figure4]{\ref*{fig:figure4}(h)}} predicts it just above $\Delta\varphi_\text{ext}=0.4$. We expect this small flux offset to be caused by small inaccuracies in the circuit's capacitances that were simulated and fitted in \mbox{Appendix \ref{app:array_modes}}. Finally, the regions of increased MIST identified by the simulations with $\ket{q}=\ket{1}$ and $\Delta\varphi_\text{ext}>0.25$ do not appear in the experimental results as a result of the decreased $T_1$ time at these flux points.

A key observation is that the results in \mbox{Figs. \hyperref[fig:figure4]{\ref*{fig:figure4}(d),(h)}} show that, at sweet spot, the readout performance of this fluxonium is limited by MIST induced by the first array mode. As noted in \mbox{Ref. \cite{fx_ro_array_modes}}, symmetry should prevent the odd array modes from coupling to the fluxonium mode. In this implementation, however, this symmetry protection is lifted by a small imbalance in the effective charging energy of the upper and lower islands of the fluxonium circuit which skews the fluxonium mode. Improved circuit design could restore this balance and the symmetry protection, providing a straightforward approach for suppressing MIST induced by the odd array modes.  

As a result of the large variation of the resonator frequency relative to its linewidth, the readout drive tone may not populate the resonator at every flux point. Additionally, the relatively short 1 $\mu$s $\sim$ $1.5/\kappa$ duration of the readout drive tone does not drive the resonator to an equilibrium, such that the average photon number occupation of the resonator can not be estimated simply as $\langle N_\text{R}^{\ket{q}} \rangle = (\Omega/2)^2/(\Delta^2 + (\kappa/2)^2)$, where $\Delta=\omega_d - \omega_\text{R}$ is the detuning between the drive frequency and the resonator frequency. Furthermore, as exemplified in Fig. \hyperref[fig:figure1]{\ref*{fig:figure1}(c)}, there are flux points where the resonator is strongly non-linear, and the photon-number-dependence of $\Delta$ must be taken into account. To obtain an estimate of the photon number occupation of the resonator, we numerically compute the classical resonator field $\alpha(t)$ by solving
\begin{equation}
    \frac{\partial \alpha(t)}{\partial t} = -\left\{ i\left[\omega_d - \omega_\text{R}^{\ket{q}}\left(\vert \alpha(t) \vert^2\right)\right] + \frac{\kappa}{2}\right\} \alpha(t) + \frac{\Omega}{2}.
\end{equation}
\noindent Here, we have made the semi-classical approximation and the rotating-wave approximation. Furthermore, the resonator frequency depends explicitly on the fluxonium state $\ket{q}$ and the average number of photons $\langle N_\text{R} \rangle=|\alpha(t)|^2$, and this dependence is found numerically with branch analysis. For each flux point and drive strength $\Omega$, we numerically solve $\alpha(t)$ and report the maximum average photon number occupation $|\alpha(t)|^2$ during the pulse in \mbox{Figs. \hyperref[fig:figure4]{\ref*{fig:figure4}(j),(k)}}. In general, when comparing Figs. \hyperref[fig:figure4]{\ref*{fig:figure4}(b)-(e)} to Figs. \hyperref[fig:figure4]{\ref*{fig:figure4}(f)-(i)}, we find that the dynamical semi-classical simulations slightly underestimate the resonator population. We ascribe this to the small resonator linewidth that may cause the resonator to not be in an equilibrium, resulting in an inaccurate conversion of the room-temperature drive strength to the drive strength at the device.

\section{Discussion}
In this work, we have experimentally investigated and modeled MIST over the full external flux range of a fluxonium qubit. The experimental results identified eleven regions with increased MIST, and for all of the regions the origin was explained using branch analysis techniques. Specifically, six were explained by avoided crossings in the dressed Hilbert space of the resonator and fluxonium, and five by avoided crossing in the Hilbert space that additionally includes the lowest two array modes. The results further highlight the role of the fluxonium's array modes in the readout process, and show that these effects are well captured by existing models, such that they may inform the design of future fluxonium quantum processors. For instance, for the fluxonium studied in this work, the readout performance at sweet spot could be improved significantly by balancing the self-capacitances of the upper and lower nodes of the circuit to suppress the coupling to the lowest array mode.

\section*{Acknowledgments}
The authors acknowledge the use of computational resources of the DelftBlue supercomputer, provided by Delft High Performance Computing Centre \cite{DHPC2024}. The authors acknowledge support from the Dutch Research Council (NWO). Additionally, J.H. acknowledges support from NWO Open Competition Science M. and E.Y.H. acknowledges support from Holland High Tech (TKI).

M.F.S.Z., J.H. and E.Y.H performed the experiments, and M.F.S.Z. analyzed the data and performed the numerical simulations. J.H., F.Y. and S.S. designed and fabricated the device. C.K.A. supervised the work. M.F.S.Z. wrote the manuscript with input from all authors.

\section*{Data Availability}
All numerical and experimental data is available through Ref. \cite{data_repository} and the code used for the numerical simulations and data processing is available through \mbox{Ref. \cite{github_repo}}.

\appendix
\section{Extended Data and Measurement Setup}
\label{app:extended_data}
\begin{figure}[t]
\centering
\includegraphics[width=\linewidth]{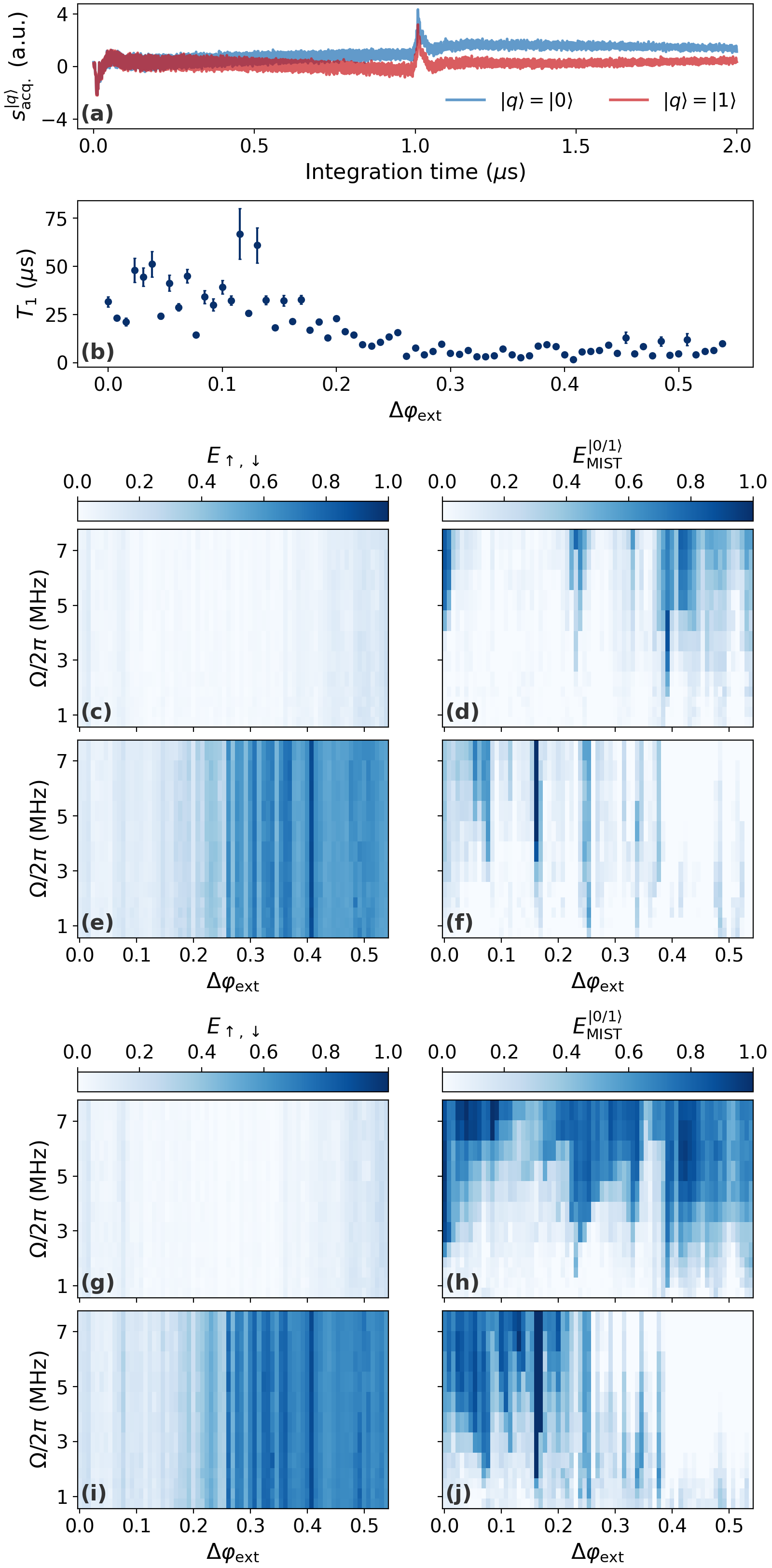}\\
\caption{\label{fig:extended_data} (a) Demodulated, acquired traces along the axis of maximum signal difference as a function of the integration time. (b) $T_1$ as a function of the flux pulse amplitude $\Delta\varphi_\text{ext}$. (c)-(f), (g)-(j) Experimental results analogous to Figs. \hyperref[fig:figure4]{\ref*{fig:figure4}(b)-(e)} in the main text, but with a 0.5 $\mu$s and 2 $\mu$s readout drive pulse duration, respectively. Panels (c)-(f) and (g)-(j) show the measured values for $E_\uparrow$, $E_\text{MIST}^{\ket{0}}$, $E_\downarrow$ and $E_\text{MIST}^{\ket{1}}$, respectively.}
\end{figure}

For each point in \mbox{Figs. \hyperref[fig:figure3]{\ref*{fig:figure3}(b),(c)}} we perform trace acquisitions to compute the optimal integration weights. Specifically, we obtain the demodulated, complex-valued, qubit-state-dependent acquired traces $s_\text{acq}^{\ket{q}}$, from which the optimal weights are calculated as $w = s_\text{acq}^{\ket{1}} - s_\text{acq}^{\ket{0}}$. In \mbox{Fig. \hyperref[fig:extended_data]{\ref*{fig:extended_data}(a)}} we plot the acquired traces along the axis of maximum signal difference for the readout parameters used in this work, of which the single-shot readout is shown in Fig. \hyperref[fig:figure3]{\ref*{fig:figure3}(d)}.

Figure \hyperref[fig:extended_data]{\ref*{fig:extended_data}(b)} shows the measured $T_1$ of the qubit as a function of the flux pulse amplitude $\Delta\varphi_\text{ext}$. The qubit is prepared in the $\ket{1}$ state at sweet spot, and tuned to a specified external flux with a fast flux pulse to characterize $T_1$ at that external flux value. For $\Delta\varphi_\text{ext}>0.25$, the $T_1$ decreases to only a few microseconds, which strongly limits the detection of MIST at these external flux values when the fluxonium is prepared in $\ket{1}$.

Figures \hyperref[fig:figure4]{\ref*{fig:figure4}(b)-(e)} in the main text showed the experimental results for a 1 $\mu$s readout pulse duration. To verify the consistency of the experiment, the experiments were repeated with a 0.5 $\mu$s and 2 $\mu$s readout pulse duration, of which the results are shown in Figs. \hyperref[fig:extended_data]{\ref*{fig:extended_data}(c)-(f)} and Figs. \hyperref[fig:extended_data]{\ref*{fig:extended_data}(g)-(j)}, respectively. We find that the same regions with increased MIST appear in these experiments, and that they show excellent resemblance with the results shown in the main text. 

\begin{figure}[b]
\centering
\includegraphics[width=\linewidth]{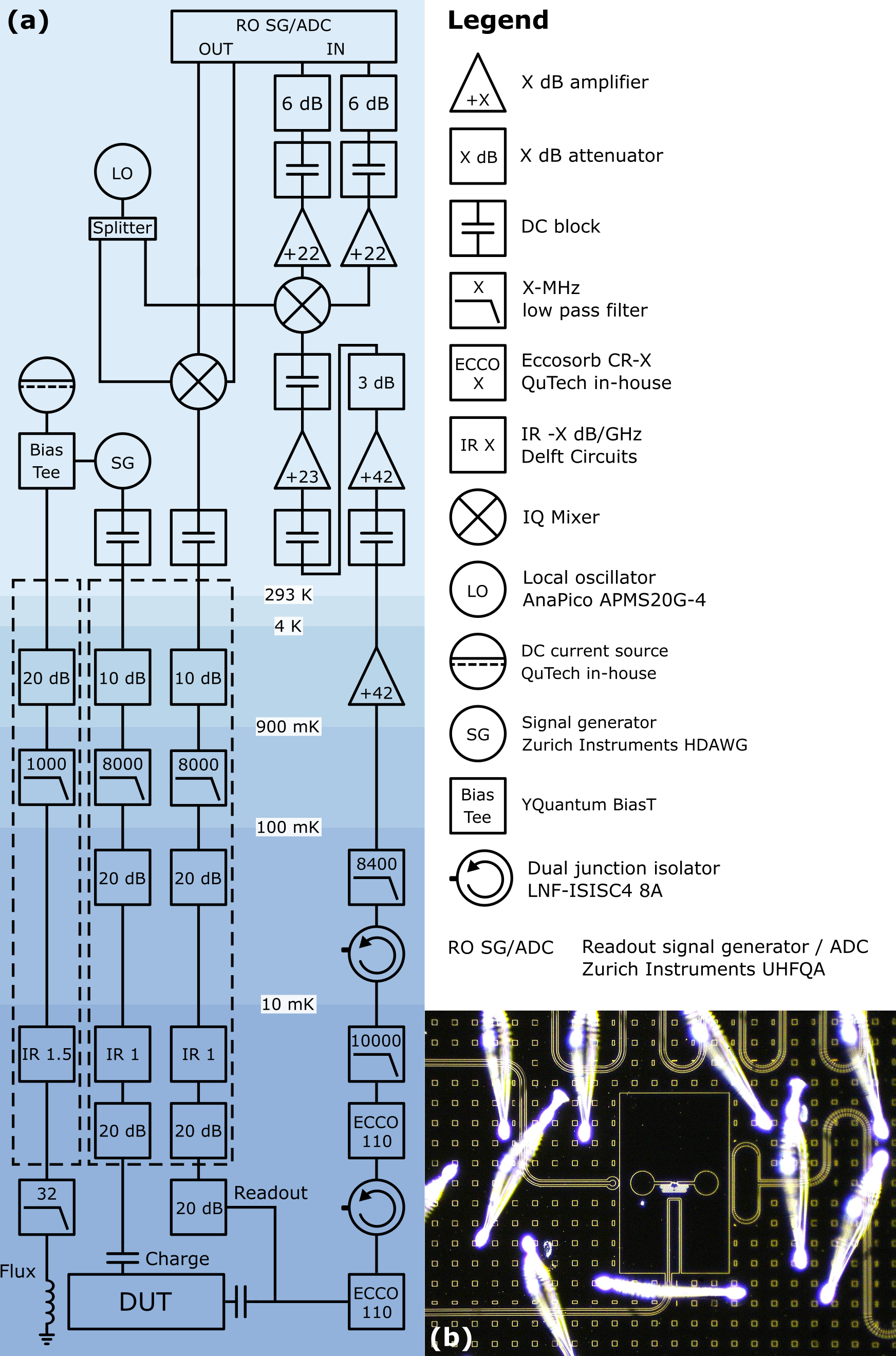}\\
\caption{\label{fig:measurement_setup} (a) Schematic of the electronic setup. (b) Optical microscope image of the device. }
\end{figure}

Figure \hyperref[fig:measurement_setup]{\ref*{fig:measurement_setup}(a)} shows the electronic setup of the device. The device, of which an optical microscope image is shown in Fig. \hyperref[fig:measurement_setup]{\ref*{fig:measurement_setup}(b)} and which has a similar design and fabrication process as Ref. \cite{pyepr}, is mounted in a 20-port YQuantum sample box. The fluxonium has a dedicated flux line, charge line, and readout line. As the frequency of the fluxonium is within the bandwidth of the signal generator (SG), no further upconversion of the charge drive pulses is required. For the flux line, a static DC current generated by an in-house made DC current generator and an RF signal for fast flux pulses generated by the SG are combined in a bias tee at room temperature. Prior to the experiments, the long-time-scale flux distortions introduced by the bias tee and other electronic components were corrected with Ramsey-type experiments. All readout pulses are generated and analyzed by a Zurich Instruments UHFQA. The electronics inside the dashed rectangles are implemented in Delft Circuits flex wiring. 

\section{Array Modes}
\label{app:array_modes}
\begin{figure}[t]
\centering
\includegraphics[width=\linewidth]{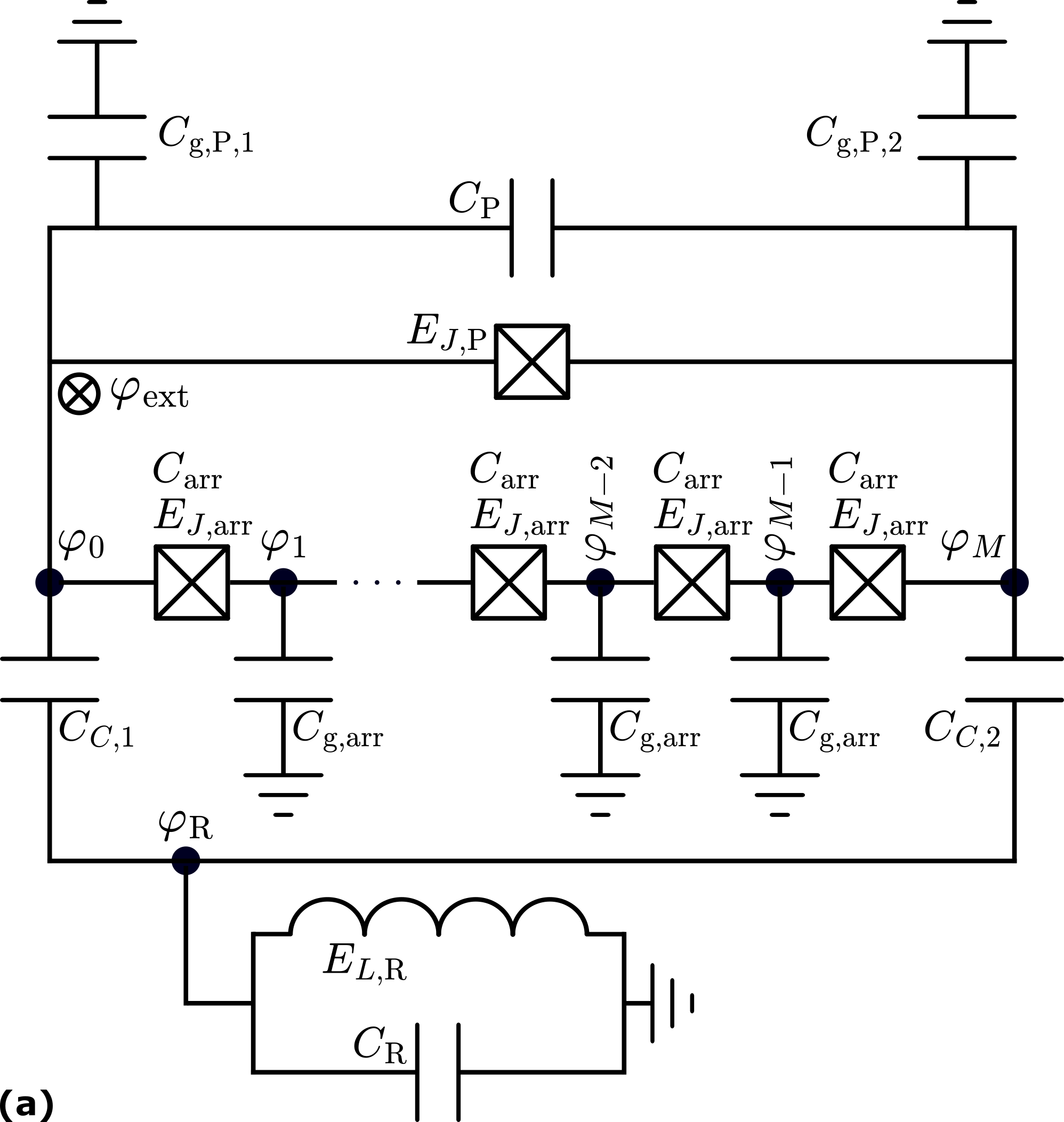}\\
\vspace{.5em}
\includegraphics[width=\linewidth]{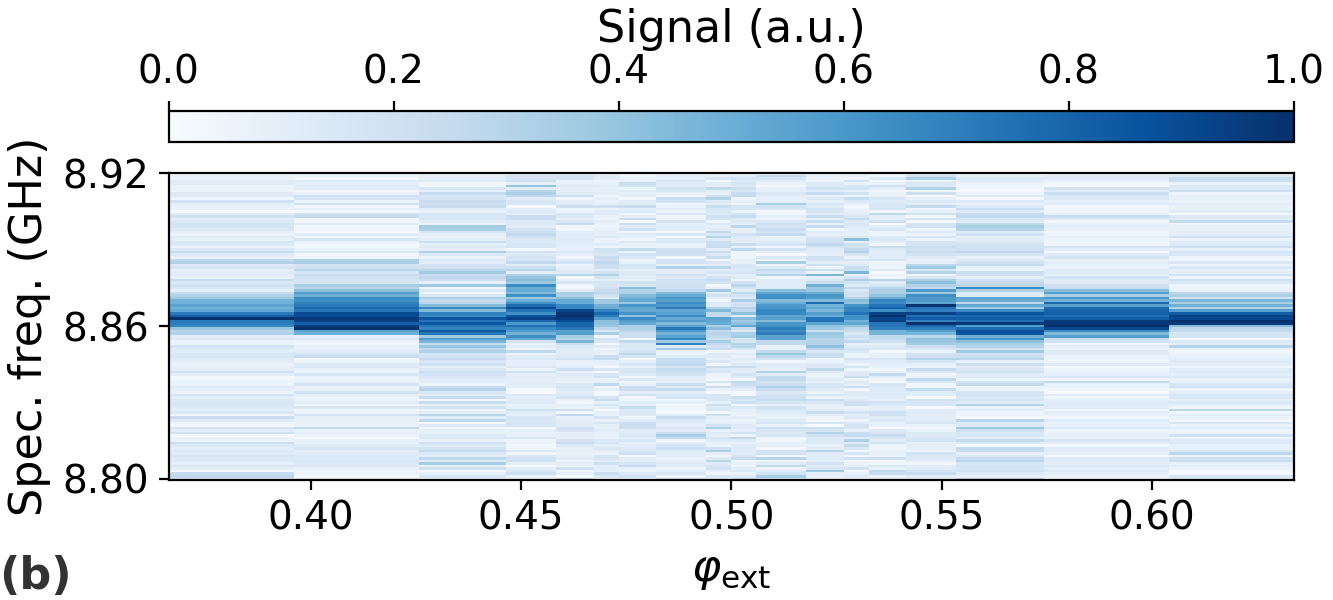}
\caption{\label{fig:FX-array} (a) Circuit diagram of a floating fluxonium with the individual Josephson junctions that make-up the array, and with the capacitively coupled readout resonator. (b) Two-tone spectroscopy of the first array mode versus the external flux $\varphi_\text{ext}$ and the spectroscopy frequency (spec. freq.).}
\end{figure}

The linear inductor of the fluxonium used in this work is realized by an array of $M=150$ Josephson junctions. The full circuit diagram of a floating fluxonium capacitively coupled to a readout resonator with the individual treatment of the JJs in the array is shown in \mbox{Fig. \hyperref[fig:FX-array]{\ref*{fig:FX-array}(a)}}. Naturally, this circuit consists of $M+1$ modes: a fluxonium mode, a resonator mode, and $M-1$ array modes. As shown in \mbox{Fig. \hyperref[fig:FX-array]{\ref*{fig:FX-array}(b)}}, a two-tone spectroscopy versus $\varphi_\text{ext}$ experiment reveals the frequency of the first array mode at $\omega_{\mu_1}/2\pi = 8.86$ GHz. To determine the frequency of the second array mode and the coupling strength of the array modes to the fluxonium and resonator, the capacitances shown in \mbox{Fig. \hyperref[fig:FX-array]{\ref*{fig:FX-array}(a)}} need to be determined. The capacitances to ground of the lower and upper nodes $C_{\text{g,P},1}=27.518$ fF and $C_{\text{g,P},2}=31.272$ fF, the resonator capacitance $C_\text{R}=679.220$ fF and the coupling capacitance \mbox{$C_{C,2}=0.839$ fF} are obtained from the design \mbox{simulations \cite{pyepr}}. We assume a capacitance of 50 fF/$\mu$m$^2$ for the JJ capacitances to find $C_\text{arr}=16.3$ fF \cite{pyepr, JJ_capacitance}. The remaining phase-slip junction capacitance \mbox{$C_\text{P} = 5.056$ fF}, coupling capacitance $C_{C,1}=5.583$ fF, ground capacitance of the array nodes \mbox{$C_\text{g,arr}=0.05637$ fF} and the frequency of the resonator are fitted to the experimentally obtained values of $E_{C,\text{F}}$, $\omega_{\mu_1}$ and the qubit-state-dependent vacuum resonator frequencies shown in \mbox{Fig. \hyperref[fig:figure1]{\ref*{fig:figure1}(c)}}. Finally, the approach detailed in \mbox{Refs. \cite{fx_ro_array_modes, mist_fx_chapple}} is used to compute the frequency of the second array mode and the coupling strengths of the array modes to the fluxonium and resonator reported in the bottom half of \mbox{Table \ref{tab:eparams}}.

\bibliography{apssamp}

\end{document}